\documentclass[10pt]{article}
\begin{document}

\title{{\bf Anomalies, Horizons and Hawking radiation}}
\author{
{Sunandan Gangopadhyay$^{}$\thanks{sunandan.gangopadhyay@gmail.com,
sunandan@sun.ac.za}}\\
National Institute for Theoretical Physics (NIThep),\\ Stellenbosch Institute
for Advanced Study (STIAS), 7600 Stellenbosch, South Africa\\[0.3cm]
}

\maketitle

\begin{abstract}
\noindent 
Hawking radiation is obtained from the Reissner-Nordstr\"{o}m blackhole 
with a global monopole and the Garfinkle-Horowitz-Strominger blackhole 
falling in the class of the most general spherically symmetric blackholes
$(\sqrt{-g}\neq1)$, using only chiral anomaly near the event 
horizon and covariant boundary condition at the event horizon. The approach
differs from the anomaly cancellation approach since apart from the
covariant boundary condition, the chiral anomaly near 
the horizon is the only input to derive the Hawking flux. 
\\[0.3cm]
\\[0.3cm]


\end{abstract}

\noindent {\it{Introduction :}}

\noindent On quantising matter fields in a background
black hole spacetime, Hawking radiation is obtained, a result
that cannot be obtained classically. Ever since Hawking's original
paper \cite{Hawking:sw, Hawking:rv}, there have been 
several derivations \cite{fulling}, \cite{parikh} 
and all of them take the quantum
effect of fields in blackhole backgrounds into account in various ways. 

\noindent A few years back, Robinson and Wilczek \cite{rw}
advanced a new approach known as the anomaly 
cancellation approach to derive Hawking
radiation from a Schwarzschild--type black hole, where 
diffeomorphism symmetry plays a significant role.
The crucial observation that they make is that, near the horizon,
black hole dynamics is effectively described by a two dimensional
chiral theory that breaks diffeomorphism symmetry. Hence, this
chiral theory is anomalous. Requiring that
the complete theory with contributions near the horizon, outside
the horizon and inside the horizon be anomaly free, a condition
is obtained from which the Hawking flux is identified.
The method was soon extended to the case of 
charged blackholes \cite{iso}. Further applications of this
approach using the consistent form of the chiral anomaly 
and later with the covariant form of the chiral anomaly \cite{bert} 
may be found in \cite{Muratasoda1}-\cite{sgsingle}. 

\noindent However, certain problems still persist in this derivation.
The universality of Hawking radiation requires that the flux gets 
determined only from information at the horizon. The point to be
noted is that, apart from the anomalous Ward identity at the horizon,
the normal Ward identity outside the horizon is also required. Furthermore,
it is also necessary to interpret an additional Wess-Zumino term as a 
contribution from the (classically irrelevant) ingoing modes. 
The question is whether it is possible to derive the flux just from
the information of the chiral anomaly at the horizon. The fact that this
is indeed so have been shown in \cite{rbsingle} 
using covariant chiral anomalies.
It has been observed from the structure of the anomaly and imposing asymptotic
flatness of the metric that the anomaly vanishes in the asymptotic limit
($r\rightarrow\infty$ limit). Hence, the Hawking flux, which is measured
at infinity, is given  by the $r\rightarrow\infty$ limit of the $r$--$t$
component of the energy-momentum tensor near the horizon since this
would correspond to the anomaly free expression associated with the
$r$--$t$ component of the energy-momentum tensor outside the horizon.
The Hawking flux is therefore obtained solely from the information near the
horizon bypassing all the problems encountered in the anomaly cancellation
approach. 

\noindent In this paper, we extend this method for the most general
spherically symmetric black hole spacetime ($\sqrt{-g}\neq1$) which may
not be asymptotically flat, e.g. the Reissner-Nordstr\"{o}m black hole 
with a global monopole \cite{vil}. Further, we derive the Hawking
flux using both forms of the chiral anomaly, namely the consistent and the
covariant. However, in case of the consistent anomaly, a little bit of more
work is required since one needs to add a local counter term 
\cite{bardeenzum}, \cite{bert1} to the
consistent current/energy momentum tensor to obtain the covariant 
current/energy momentum tensor which is required for imposing the
covariant boundary condition at the horizon.

\noindent Examples of the Reissner-Nordstr\"{o}m black hole 
with a global monopole and the  Garfinkle-Horowitz-Strominger (GHS)
black hole \cite{ghs} are finally discussed.\\


\noindent {\it{New approach of deriving the Hawking radiation :}}\\

\noindent We start with the most general spherically symmetric 
black hole spacetime $(\sqrt{-g}\neq1)$ given by
\begin{eqnarray}
ds^2 &=& f(r)dt^2 -h(r)^{-1}dr^2 +r^2d\Omega^2~.
\label{spher_metric}
\end{eqnarray}
With the aid of dimensional reduction procedure one can effectively describe
a theory with a metric given by the ``$r-t$" sector of the full 
spacetime metric (\ref{spher_metric}) near the horizon \cite{rw}.

\noindent Now we divide the spacetime into two regions and concentrate
only in the near horizon region to discuss the gauge/gravitational 
anomalies separately. We shall first derive the Hawking flux using 
the consistent chiral anomaly and then with the covariant chiral anomaly.\\

\noindent {\it{Consistent Gauge anomaly :}}\\

\noindent Near the horizon there are only outgoing (right-handed) fields
and the current becomes anomalous. The consistent 
form of $d=2$ Abelian anomaly satisfies \cite{iso} 
\begin{eqnarray}
\nabla_{\mu}J^{\mu}_{(H)}&=& -\frac{e^2}{4\pi}
\bar\epsilon^{\rho\sigma}\partial_{\rho}A_{\sigma}= 
\frac{e^2}{4\pi\sqrt{-g}}\partial_{r}A_{t}
\label{10.1}  
\end{eqnarray}  
where, $\bar\epsilon^{\mu\nu}=
\epsilon^{\mu\nu}/\sqrt{-g}$
and $\bar\epsilon_{\mu\nu}=
\sqrt{-g}\epsilon_{\mu\nu}$ are two dimensional antisymmetric tensors 
for the upper and lower cases with $\epsilon^{tr}=\epsilon_{rt}=1$.

\noindent It is easy to see that the above anomaly equation (\ref{10.1})
leads to the following differential equation
\begin{equation}
\partial_{r}\left(\sqrt{-g}J^{r}_{(H)}\right) 
= \frac{e^2}{4\pi}\partial_{r}A_{t}~.
\label{10.2}
\end{equation}      
Solving (\ref{10.2}) in the region near the horizon, we get
\begin{eqnarray}
J^{r}_{(H)}(r) &=& \frac{1}{\sqrt{-g}}
\left(a_{H}+\frac{e^2}{4\pi}\int_{r_H}^{r}
\partial_{r}A_{t}(r)\right)\nonumber\\
&=&\frac{1}{\sqrt{-g}}(a_{H} +\frac{e^2}{4\pi}
\left[A_{t}(r) - A_{t}(r_{H})\right])
\label{10.3}
\end{eqnarray}
where, $a_{H}$ is an integration constant. 
This constant $a_{H}$ gets fixed by requiring that 
the covariant current $\tilde{J}^{r}_{(H)}(r)$ 
vanishes at the horizon. To impose this boundary
condition, we recall that the covariant and consistent
currents are related by local counter terms \cite{bardeenzum}
\begin{eqnarray}
\tilde{J}^{\mu}_{(H)}(r) &=& J^{\mu}_{(H)}(r)+\frac{e^2}{4\pi\sqrt{-g}}
A_{\lambda}\epsilon^{\lambda\mu}
\label{10.4}
\end{eqnarray}
which using (\ref{10.3}) leads to
\begin{eqnarray}
\sqrt{-g}\tilde{J}^{r}_{(H)}(r) &=& a_{H} +\frac{e^2}{2\pi}
A_{t}(r) - \frac{e^2}{4\pi}A_{t}(r_{H})~.
\label{10.5}
\end{eqnarray}
Now the condition that the covariant current $\tilde{J}^{r}_{(H)}(r)$ 
vanishes at the horizon yields
\begin{eqnarray}
a_{H}=-\frac{e^2}{4\pi}A_{t}(r_{H})~.
\label{10.6}
\end{eqnarray}
The charge flux can now be obtained by taking 
the asymptotic limit of (\ref{10.3}) and is given by  
\begin{equation}
(\sqrt{-g}J^{r}_{(H)})(r\rightarrow\infty)
=-\frac{e^2}{2\pi}A_{t}(r_{H})~.
\label{10.7}
\end{equation}
This is precisely the charge flux obtained in (\cite{rw})
using the method of cancellation of consistent gauge
anomaly.\\

\noindent {\it{Consistent Gravitational anomaly :}}\\

\noindent The consistent form of $d=2$ gravitational anomaly is 
given by (\cite{rw, iso}):
\begin{equation}
\nabla_{\mu}T^{\mu}_{(H)\nu} = F_{\mu\nu}J^{\mu}_{(H)}
+A_{\nu}\nabla_{\mu}J^{\mu}_{(H)}+\mathcal A_{\nu}
\label{10.8}
\end{equation}
where, $\mathcal A_{\nu}$ is the consistent form of the gravitational
anomaly given by 
\begin{equation}
\mathcal A_{\nu}=\frac{1}{96\pi\sqrt{-g}}\epsilon^{\beta\delta}
\partial_{\delta}\partial_{\alpha}{\Gamma^{\alpha}}_{\nu\beta}~.
\label{10.9}
\end{equation}
For the $\nu=t$ component, the above equation simplifies to
\begin{eqnarray}
\nabla_{\mu}T^{\mu}_{(H)t}&=&
F_{rt}\tilde{J}^{r}_{(H)}+\mathcal A_{t}
\label{11.1}
\end{eqnarray}
where, 
\begin{eqnarray}
\mathcal A_{t}&=& \frac{1}{\sqrt{-g}}\partial_{r}N^{r}_{t}\quad;\quad
\mathcal A_{r}=0\nonumber\\
N^{r}_{t}&=&\frac{1}{192\pi}\left(hf'' + f'h'\right).
\label{110.1}
\end{eqnarray}
Equation (\ref{11.1}) finally leads to
\begin{eqnarray}
\partial_{r}\left(\sqrt{-g}T^{r}_{(H)t}\right)
&=&\sqrt{-g}\left[F_{rt}\tilde{J}^{r}_{(H)}+\mathcal A_{t}\right]\nonumber\\
&=&\sqrt{-g}F_{rt}\tilde{J}^{r}_{(H)}+\partial_{r}N^{r}_{t}\nonumber\\
&=&\frac{e^2}{2\pi}[A_{t}(r)-A_{t}(r_H)]\partial_{r}A_{t}
+\partial_{r}N^{r}_{t}~.
\label{11.2}
\end{eqnarray}
Solving the above equation, we get
\begin{eqnarray}
\sqrt{-g}T^{r}_{(H)t}
&=&d_{H}+\frac{e^2}{4\pi}[A^{2}_{t}(r)-A^{2}_{t}(r_H)]-\frac{e^2}{2\pi}
A_{t}(r_H)[A_{t}(r)-A_{t}(r_H)]\nonumber\\
&&~~~~~~~~~~~~~~~~+N^{r}_{t}(r)-N^{r}_{t}(r_{H})~.
\label{11.3}
\end{eqnarray}
where, $d_{H}$ is an integration constant. This constant 
$d_{H}$ gets fixed by requiring that 
the covariant energy momentum tensor $\tilde{T}^{r}_{(H)t}(r)$ 
vanishes at the horizon. To impose this boundary
condition, we note that the covariant and consistent
energy momentum tensors are related by local counter terms 
(see Appendix for a derivation)
\begin{eqnarray}
\sqrt{-g}\tilde{T}^{r}_{(H)t}
&=&\sqrt{-g}T^{r}_{(H)t}+\frac{h}{192\pi f}(ff''-2f'^2)
\label{11.4}
\end{eqnarray}
which leads to 
\begin{eqnarray}
\sqrt{-g}\tilde{T}^{r}_{(H)t}
&=&d_{H}+\frac{e^2}{4\pi}[A^{2}_{t}(r)-A^{2}_{t}(r_H)]
-\frac{e^2}{2\pi}A_{t}(r_H)[A_{t}(r)-A_{t}(r_H)]\nonumber\\
&&+N^{r}_{t}(r)-N^{r}_{t}(r_{H})+\frac{h}{192\pi f}(ff''-2f'^2)~.
\label{11.5}
\end{eqnarray}
Now the condition that the covariant current $\tilde{T}^{r}_{(H)t}(r)$ 
vanishes at the horizon yields
\begin{eqnarray}
d_{H}=\frac{1}{96\pi}f'(r_{H})h'(r_{H})~.
\label{11.6}
\end{eqnarray}
The energy flux can now be obtained by taking 
the asymptotic limit of (\ref{11.3}) and is given by  
\begin{equation}
(\sqrt{-g}T^{r}_{(H)t})(r\rightarrow\infty)
=\frac{e^2}{4\pi}A^{2}_{t}(r_{H})+ \frac{1}{192\pi}f'(r_{H})h'(r_{H})~.
\label{11.7}
\end{equation}
This is precisely the energy flux obtained in (\cite{rw})
using the method of cancellation of consistent gravitational
anomaly.\\
 

\noindent {\it{Covariant Gauge anomaly :}}\\

\noindent The covariant form of the $d=2$ gauge anomaly satisfies \cite{iso} 
\begin{eqnarray}
\nabla_{\mu}\tilde{J}^{\mu}_{(H)}=-\frac{e^2}{4\pi}
\bar\epsilon^{\rho\sigma}F_{\rho\sigma}= 
\frac{e^2}{2\pi\sqrt{-g}}\partial_{r}A_{t}~.
\label{2.3}  
\end{eqnarray}  
It is easy to see that the above anomaly equation (\ref{2.3})
leads to the following differential equation
\begin{equation}
\partial_{r}\left(\sqrt{-g}\tilde{J}^{r}_{(H)}\right) 
= \frac{e^2}{2\pi}\partial_{r}A_{t}~.
\label{2.4bb}
\end{equation}      
Solving (\ref{2.4bb}) in the region near the horizon, we get
\begin{eqnarray}
\tilde{J}^{r}_{(H)}(r) &=& \frac{1}{\sqrt{-g}}
\left(c_{H}+\frac{e^2}{2\pi}\int_{r_H}^{r}
\partial_{r}A_{t}(r)\right)\nonumber\\
&=&\frac{1}{\sqrt{-g}}(c_{H} +\frac{e^2}{2\pi}
\left[A_{t}(r) - A_{t}(r_{H})\right])
\label{2.6}
\end{eqnarray}
where, $c_{H}$ is an integration constant. 
The constant $c_{H}$ vanishes by requiring that 
the covariant current $\tilde{J}^{r}_{(H)}(r)$ 
vanishes at the horizon. The charge flux
can now be obtained by taking the asymptotic limit of the above equation
and is given by  
\begin{equation}
(\sqrt{-g}\tilde{J}^{r}_{(H)})(r\rightarrow\infty)
=-\frac{e^2}{2\pi}A_{t}(r_{H})~. 
\label{2.10}
\end{equation}
The above result agrees with (\ref{10.7}) and is precisely the 
charge flux obtained in (\cite{rb, sgsk, sgsingle})
using the method of cancellation of covariant gauge
anomaly. \\ 

\noindent {\it{Covariant Gravitational anomaly :}}\\

\noindent The covariant form of $d=2$ 
gravitational anomaly is given by (\cite{rw, iso}):
\begin{equation}
\nabla_{\mu}\tilde{T}^{\mu}_{(H)\nu} = F_{\mu\nu}\tilde{J}^{\mu}_{(H)}
+\frac{1}{96\pi}
\bar\epsilon_{\nu\mu}\partial^{\mu}R 
= F_{\mu\nu}\tilde{J}^{\mu}_{(H)}+\tilde{\mathcal A}_{\nu}~.
\label{cov}
\end{equation}
It is easy to check that for the metric (\ref{spher_metric}), 
the two dimensional Ricci scalar $R$ is given by 
\begin{eqnarray}
R=\frac{h~f^{''}}{f}+\frac{f^{'}h^{'}}{2f}
-\frac{f^{'2}h}{2f^2}
\label{ricci}
\end{eqnarray}
and the anomaly is purely timelike with
\begin{eqnarray}
\tilde{\mathcal A_{r}} &=& 0\nonumber\\
\tilde{\mathcal A_{t}} &=& \frac{1}{\sqrt{-g}}\partial_{r}\tilde{N}^{r}_{t}
\label{8}
\end{eqnarray}
where,
\begin{equation}
\tilde{N}^{r}_{t} = \frac{1}{96\pi}\left(hf'' + 
\frac{f'h'}{2} - \frac{f'^{2}h}{f}\right).
\label{9}
\end{equation}
We now solve the anomaly equation (\ref{cov}) 
for the $\nu=t$ component and this leads 
to the following differential equation for the
$r$--$t$ component of the near horizon covariant energy-momentum tensor
\begin{eqnarray}
\partial_{r}\left(\sqrt{-g}\tilde{T}^{r}_{(H)t}\right)&=& 
\sqrt{-g}F_{rt}\tilde{J}^{r}_{(H)}(r)
+\partial_{r}\tilde{N}^{r}_{t}(r)\nonumber\\
&=&(c_{H} +\frac{e^2}{2\pi}
\left[A_{t}(r) - A_{t}(r_{H})\right])\partial_{r}A_{t}(r)
+\partial_{r}\tilde{N}^{r}_{t}(r)\nonumber\\
&=&\partial_{r}\left(\frac{e^2}{2\pi}\left[\frac{1}{2}A^2_{t}(r)-
A_{t}(r_{H})A_{t}(r)\right]+\tilde{N}^{r}_{t}(r)\right)
\label{900}
\end{eqnarray}      
where we have used (\ref{2.6}) in the 
second line and set $c_{H}=0$ in the last line of the above equation.
Integration of the above equation leads to
\begin{eqnarray}
\tilde{T}^{r}_{(H)t}(r)&=&\frac{1}{\sqrt{-g}}\left(b_{H}+\int_{r_H}^{r}
\partial_{r}\left(\frac{e^2}{2\pi}
\left[\frac{1}{2}A^{2}_{t}(r) - A_{t}(r_{H})A_{t}(r)\right]
+\tilde{N}^{r}_{t}(r)\right)
\right)
\nonumber\\
&=& \frac{1}{\sqrt{-g}}
\left(b_{H} +\frac{e^2}{4\pi}[A^2_{t}(r)+A^2_{t}(r_H)]
-\frac{e^2}{2\pi}A_{t}(r_{H})A_{t}(r)+ 
\tilde{N}^{r}_{t}(r) - \tilde{N}^{r}_{t}(r_{H})\right)
\label{10}
\end{eqnarray}
where, $b_{H}$ is an integration constant. 
The integration constant $b_{H}$ can be fixed by imposing 
that the covariant energy-momentum tensor vanishes at the horizon. 
From (\ref{10}), this gives $b_{H}=0$.
Hence the total flux of the energy-momentum tensor is given by
\begin{eqnarray}
(\sqrt{-g}\tilde{T}^{r}_{(H)t})(r\rightarrow\infty)
&=&\frac{e^2}{4\pi}A^{2}_{t}(r_H)-\tilde{N}^{r}_{t}(r_{H})\nonumber\\
&=&\frac{e^2}{4\pi}A^{2}_{t}(r_H)+\frac{1}{192\pi} f'(r_{H})h'(r_{H})~.
\label{flux}
\end{eqnarray}\\
The above result agrees with (\ref{11.7}) and is 
precisely the Hawking flux obtained in (\cite{rb, sgsk, sgsingle})
using the method of cancellation of covariant gravitational anomaly. \\

\noindent {Examples :}\\

\noindent {\it{Hawking radiation from Reissner-Nordstr\"{o}m blackhole 
with a global monopole :}}\\

\noindent The metric of a general non-extremal 
Reissner-Nordstr\"{o}m blackhole with a global monopole $O(3)$ is given by
\cite{vil}
\begin{equation}
ds^{2}_{string} = p(r)dt^{2} - \frac{1}{h(r)}dr^{2} - r^{2}d\Omega^2
\label{200} 
\end{equation}
where,
\begin{eqnarray}
A = \frac{q}{r}dt\quad,\quad 
p(r)= h(r) = 1 -\eta^2 -\frac{2m}{r} +\frac{q^2}{r^2} 
\label{200a}
\end{eqnarray}
with $m$ being the mass parameter of the blackhole and 
$\eta$ is related to the symmetry breaking
scale when the global monopole is formed during 
the early universe soon after the Big-Bang \cite{JP}.
The event horizon for the above blackhole is situated at   
\begin{eqnarray}
r_{H}&=&(1 -\eta^2)^{-1}\big[m +\sqrt{m^2 -(1 -\eta^2)q^2}\big]~.
\label{hor}
\end{eqnarray}
Now it has been argued in \cite{peng} that the metric (\ref{200}) 
has to be rewritten in the form (\ref{spher_metric}) with
\begin{eqnarray}
f(r) &=&(1 -\eta^2)h(r) \qquad, \qquad 
h(r) = 1 -\eta^2 -\frac{2m}{r} -\frac{q^2}{r^2}~. 
\label{newmetric}
\end{eqnarray}
in order to get the correct Hawking temperature for the metric (\ref{200})
by the anomaly cancellation approach.
We shall take this form of the metric (\ref{newmetric}) to derive the
Hawking flux. Note that the determinant of the above metric
$\sqrt{-g}\neq 1$.

\noindent Using either of the equations (\ref{10.7}, \ref{2.10}), the
charge flux is given by
\begin{equation}
(\sqrt{-g}J^{r}_{(H)})(r\rightarrow\infty)=
(\sqrt{-g}\tilde{J}^{r}_{(H)})(r\rightarrow\infty)
=-\frac{e^2 q}{2\pi r_{H}}
\label{100.7}
\end{equation}
which agrees with \cite{sgsingle}.
 
\noindent The energy flux can be obtained  by using 
either of the equations (\ref{11.7}, \ref{flux}) :
\begin{equation}
(\sqrt{-g}T^{r}_{(H)t})(r\rightarrow\infty)=
(\sqrt{-g}\tilde{T}^{r}_{(H)t})(r\rightarrow\infty)
=\frac{e^2 q^2}{4\pi r_{H}^2}+ \frac{1}{192\pi}
\frac{f'^2(r_{H})}{(1-\eta^2)}~.\\
\label{110.7}
\end{equation}
This is precisely the Hawking flux obtained in \cite{sgsingle} using the
anomaly cancellation as well as the effective action approach.\\

\noindent {\it{Garfinkle-Horowitz-Strominger blackhole :}}\\

\noindent The metric of the GHS blackhole \cite{ghs} is 
of the form (\ref{spher_metric}) where
\begin{eqnarray}
f(r) &=& \left( 1- \frac{2Me^{\phi_{0}}}{r}\right)
\left(1 - \frac{Q^{2}e^{3\phi_{0}}}{Mr}\right)^{-1}\nonumber\\
h(r) &=& \left(1-\frac{2Me^{\phi_{0}}}{r}\right)
\left(1- \frac{Q^{2}e^{3\phi_{0}}}{Mr}\right)
\label{2abcd}
\end{eqnarray}
with $\phi_{0}$ being the asymptotic 
constant value of the dilaton field.
We consider the case when $Q^{2}<2e^{-2\phi_{0}}M^{2}$
for which the above metric describes a blackhole with an event horizon 
situated at   
\begin{eqnarray}
r_{H}&=&2Me^{\phi_{0}}~.
\label{hor}
\end{eqnarray}
Here, we need to consider only the chiral gravitational anomaly
since the charge sector is absent.
The energy flux can be obtained  by using 
either of the equations (\ref{11.7}, \ref{flux}) :
\begin{equation}
(\sqrt{-g}T^{r}_{(H)t})(r\rightarrow\infty)=
(\sqrt{-g}\tilde{T}^{r}_{(H)t})(r\rightarrow\infty)
=\frac{1}{192\pi}f'(r_{H})h'(r_{H})=\frac{\pi}{12(8\pi Me^{\phi_0})^2}~.
\label{110.7}
\end{equation}
This is precisely the Hawking flux obtained in 
\cite{vagenas, sgsk, sg} using the
anomaly cancellation and effective action approach.\\

\noindent {\it{Discussions :}}

\noindent In this paper, we studied the problem of Hawking radiation 
from Reissner-Nordstr\"{o}m blackhole with a global monopole
and GHS black hole using both forms (consistent and covariant) 
of the near horizon chiral anomaly.
An important advantage of this procedure in contrast to the
anomaly cancellation technique is that the chiral anomaly near the
horizon is the only ingredient (apart from the imposition of the
covariant boundary condition near the horizon) to compute the Hawking flux.\\

\noindent {\it{Appendix :}}\\

\noindent In this appendix, we shall derive the form of the local
counter term connecting the $r$--$t$ component of the near horizon
covariant energy-momentum tensor $\tilde{T}^{r}_{(H)t}$ 
with the $r$--$t$ component of the near horizon 
consistent energy-momentum tensor $T^{r}_{(H)t}$ (\ref{11.4}).

\noindent To do this, we note that the covariant and the consistent 
energy-momentum tensors are related by local counter terms \cite{bert1} :
\begin{eqnarray}
\tilde{T}_{(H)\mu\nu} = T_{(H)\mu\nu}+P_{\mu\nu}
\label{local}    
\end{eqnarray}
where, 
\begin{eqnarray}
\nabla^{\mu}P_{\mu\nu}=-\frac{1}{96\pi}
(\bar\epsilon_{\mu\nu}\nabla^{\mu}R-\bar\epsilon^{\gamma\delta}
\partial_{\alpha}\partial_{\gamma}{\Gamma^{\alpha}}_{\delta\nu}).
\label{gradlocal}    
\end{eqnarray}  
Now for the static black hole background (\ref{spher_metric}),
the above equation (\ref{gradlocal}) simplifies to
\begin{eqnarray}
\partial_{r}\left(\sqrt{-g}{P^{r}}_{t}\right)=\frac{1}{96\pi}
\partial_{r}\left(\frac{hf''}{2}-\frac{h}{f}f'^{2}\right).
\label{gradlocal_simple}    
\end{eqnarray}  
Solving the above equation, we have
\begin{eqnarray}
\sqrt{-g}{P^{r}}_{t}=\frac{1}{96\pi}
\left(\frac{hf''}{2}-\frac{h}{f}f'^{2}\right)+C
\label{gradlocal_sol}    
\end{eqnarray}  
where, $C$ is an integration constant. To determine this constant,
we take the asymptotic limit ($r\rightarrow\infty$ limit) of the above equation.
In this limit, ${P^{r}}_{t}$ vanishes since the 
$r$--$t$ components of the near horizon
covariant and consistent energy-momentum tensors $\tilde{T}^{r}_{(H)t}$ 
and $T^{r}_{(H)t}$ coincides
with the usual anomaly free  $r$--$t$ component of the 
energy-momentum tensor outside the horizon ($T^{r}_{(O)t}$) 
in the $r\rightarrow\infty$ limit. Hence, the integration constant
$C$ vanishes since the first term on the 
right hand side of (\ref{gradlocal_sol}) vanishes in the  
$r\rightarrow\infty$ limit. This leads to
\begin{eqnarray}
{P^{r}}_{t}=\frac{h}{192\pi f\sqrt{-g}}
\left(ff''-2f'^{2}\right)
\label{gradlocal_sol_rev}    
\end{eqnarray}  
which yields the required result (\ref{11.4}) connecting the
$r$--$t$ components of the near horizon
covariant and consistent energy-momentum tensors $\tilde{T}^{r}_{(H)t}$ 
and $T^{r}_{(H)t}$.


\end{document}